\documentclass[conference]{IEEEtran}
\IEEEoverridecommandlockouts
\usepackage{cite}
\usepackage{amsmath,amssymb,amsfonts}
\usepackage{algorithmic}
\usepackage{graphicx}
\usepackage{textcomp}
\usepackage[table]{xcolor}
\usepackage[cmintegrals]{newtxmath}
\usepackage{bm}
\usepackage{array}
\usepackage{multirow}
\usepackage{enumitem}
\usepackage{dblfloatfix}
\usepackage{fixltx2e}
\usepackage{booktabs}
\usepackage{subcaption}
\usepackage{colortbl}

\definecolor{lightgray}{gray}{0.95}

\newcolumntype{L}{>{\raggedright\arraybackslash}m{3.1 cm}}
\newcolumntype{C}{>{\centering\arraybackslash}m{3.1 cm}}
\newcolumntype{K}{>{\raggedright\arraybackslash}m{2.2 cm}}
\newcolumntype{d}{>{\centering\arraybackslash}m{1.4 cm}}
\newcolumntype{D}{>{\centering\arraybackslash}m{2.1 cm}}
\newcolumntype{k}{>{\raggedleft\arraybackslash}m{1.8 cm}}
\newcolumntype{A}{>{\centering\arraybackslash}m{1.8 cm}}

\newcolumntype{V}{>{\raggedright\arraybackslash}m{4 cm}}
\newcolumntype{E}{>{\raggedright\arraybackslash}m{1 cm}}
\newcolumntype{P}{>{\centering\arraybackslash}m{1.5 cm}}
\newcolumntype{x}{>{\centering\arraybackslash}m{0.75 cm}}
\newcolumntype{z}{>{\centering\arraybackslash}m{0.45 cm}}
\newcolumntype{e}{>{\centering\arraybackslash}m{1.0 cm}}

\newcolumntype{?}{!{\vrule width 1.2pt}}

\def\BibTeX{{\rm B\kern-.05em{\sc i\kern-.025em b}\kern-.08em
    T\kern-.1667em\lower.7ex\hbox{E}\kern-.125emX}}

\usepackage{fancyhdr}
\fancypagestyle{firststyle}{\fancyhf{}
	\fancyhead[L]{\small Dipankar Shakya, Mingjun Ying, and Theodore S. Rappaport, ``Angular Spread Statistics for 6.75 GHz FR1(C) and 16.95 GHz FR3 Mid-Band Frequencies in an Indoor Hotspot Environment",\textit{ (Submitted) IEEE Wireless Communications and Networking Conference, }Milan, Mar. 2025, pp. 1--6.}
}

\begin{document}

\title{Angular Spread Statistics for 6.75 GHz FR1(C) and 16.95 GHz FR3 Mid-Band Frequencies in an Indoor Hotspot Environment}

\author{
\IEEEauthorblockN{Dipankar Shakya, Mingjun Ying, and Theodore S. Rappaport}
\IEEEauthorblockA{NYU WIRELESS, Tandon School of Engineering, New York University, Brooklyn, NY, 11201\\
 \{dshakya, yingmingjun, tsr\}@nyu.edu}
}

\maketitle

\thispagestyle{firststyle}

\begin{abstract}
We present detailed multipath propagation spatial statistics for next-generation wireless systems operating at lower and upper mid-band frequencies spanning 6--24 GHz. The large-scale spatial characteristics of the wireless channel include Azimuth angular Spread of Departure (ASD) and Zenith angular Spread of Departure (ZSD) of multipath components (MPC) from a transmitter and the Azimuth angular Spread of Arrival (ASA) and Zenith angular Spread of Arrival (ZSA) at a receiver. The angular statistics calculated from measurements were compared with industry-standard 3GPP models, and ASD and ASA values were found to be in close agreement at both 6.75 GHz and 16.95 GHz. Measured LOS ASD was found larger than 3GPP ASD indicating more diverse MPC departure directions in the azimuth. ZSA and ZSD were observed smaller than the 3GPP modeling results as most multipath arrivals and departures during measurements were recorded at the boresight antenna elevation. The wide angular spreads indicate a multipath-rich spatial propagation at 6.75 GHz and 16.95 GHz, showing greater promise for the implementation of MIMO beamforming systems in the mid-band spectrum. 

\end{abstract}


\section{Introduction}
\label{sec:intro}

The progression towards 6G wireless communications has generated significant interest in the 6 to 24 GHz spectrum, as these frequency bands offer a promising balance between coverage and capacity. Regulatory bodies such as the ITU, NTIA, and FCC have emphasized the strategic importance of these bands, particularly the 7.125-8.4 GHz, 4.40-4.80 GHz, and 14.8-15.35 GHz segments, for future wireless networks \cite{NTIA2024,FCC2019ET}. 
As the industry gears up for cellular deployments in these mid-band frequencies, understanding the propagation characteristics and angular spread statistics in indoor and outdoor environments is crucial for effective network planning and optimization.

Angular spread (AS) is a crucial parameter in wireless communication, significantly influencing key performance metrics such as beamforming effectiveness, channel hardening, spatial correlation, and channel estimation accuracy. In beamforming systems, narrower AS, typically observed at higher frequencies in the mmWave and sub-THz bands, enhance beamforming performance by focusing the signal within the narrow beams. Conversely, broader spreads at lower frequencies with omnidirectional (omni) antennas, such as 3.5 GHz, introduce challenges due to increased signal scattering, which can degrade the signal-to-interference-plus-noise ratio (SINR) \cite{Li2018TWC, Kaya2016Globecom}. Research has shown that AS has a non-monotonic effect on achievable rates in MIMO systems, particularly when transmit antenna array sizes are moderate between 20 and 50 antennas \cite{rupasinghe2018impact}. Pilot contamination is another aspect of MIMO communications where wider AS is known to degrade the channel estimation accuracy requiring longer pilot lengths \cite{alshammari2016impact}. In dense urban environments, AS plays a significant role in determining the power distribution and spectral efficiency for radio links \cite{kelner2019evaluation}. Especially considering higher frequencies in upper mid-band and mmWave spectrum, wider half-power beamwidth (HPBW) is shown to be desirable for larger AS when accounting for beam misalignment in beamforming systems\cite{kutty2019impact}.  

Moreover, AS affects the spatial correlation in antenna arrays having closely spaced elements, with lower AS leading to higher spatial correlation and reduced scattering of MPC.  This increased correlation due to lower AS can improve channel estimation accuracy, however, can negatively impact the performance of MIMO systems by reducing the effectiveness of spatial multiplexing and diversity techniques \cite{saeed2009impact, alshammari2016impact}. Additionally, the frequency-dependent nature of AS, where higher frequencies are associated with smaller AS due to increased directionality and reduced scattering, is a critical factor as wireless communications advance towards upper mid-band, mmWave, and THz frequencies. Understanding this behavior is essential for enhancing channel estimation accuracy and optimizing network performance in these higher frequency bands \cite{kelner2018comparison, you2022exploiting, wang2017sparse}.

AS behavior is of considerable importance for Air-to-Ground (A2G) links for 6G communications. Studies on A2G channels reveal that AS varies with UAV altitude, where azimuth spread of arrival (ASA) decreases and elevation/zenith spread of arrival (ZSA) increases. These findings highlight the unique characteristics of A2G channels, particularly the significant impact of ground reflections, which are crucial for designing reliable 6G networks involving UAV communications \cite{wang2019angular}. 

Despite such broad implications, there is a notable scarcity of empirical data on the propagation characteristics within the frequency ranges of 6-24 GHz. Previous studies conducted in environments such as office corridors and university hallways have explored path loss exponents and RMS delay spreads at various frequencies, but comprehensive indoor measurements at FR1(C) and FR3 bands remain limited. This paper presents spatial statistics for indoor hotspot (InH) environments, evaluated from an extensive measurement campaign encompassing indoor, factory, and outdoor environments at 6.75 GHz (FR1(C)) and 16.95 GHz (FR3) using a 1 GHz bandwidth sliding correlation channel sounder. Conducted at the NYU WIRELESS Research Center, the indoor measurements provide detailed insights into the spatial propagation behavior and angular spread characteristics, encompassing both line-of-sight (LOS) and non-LOS (NLOS) scenarios. Over 30,000 power delay profiles (PDPs) were collected to analyze angular statistics, offering unique propagation insights at these frequencies and suggesting potential revisions to existing models.

The organization of this paper is as follows: Section \ref{sec:system_and_scenario} describes the channel sounding system and the indoor hotspot scenario. Section \ref{sxn} details the angular statistical channel model at 6.75 and 16.95 GHz. Section \ref{3GPP} highlights the 3GPP AS models and Section \ref{diss_comp_3GPP} compares NYU measurement results with 3GPP models. The paper concludes with a summary of findings and implications for future wireless communication systems.

\section{Measurement System, Environment, and Procedures}
\label{sec:system_and_scenario}

\subsection{Channel Sounding System}
Measurements were conducted with a wideband sliding correlation channel sounder at 6.75 and 16.95 GHz. Key system features are described in \cite{Shakya2024ojcoms,Shakya2024gc, Shakya2024gc2} and include:
\begin{itemize}
    \item 500 Mcps PN sequence sliding correlation baseband with 1 GHz RF bandwidth for high temporal resolution of MPC delays (1 ns)
    \item Dual-band co-located RF front-end modules for efficient frequency switching. One band is active at a time, while the other front-end module is powered off.
    \item 31 dBm EIRP for adequate coverage, while staying within FCC licensed limits.
    \item Directional horn antennas with 15 dBi (6.75 GHz) and 20 dBi (16.95 GHz) gain mounted on mechanically rotatable gimbals with a one-degree spatial resolution for directional measurements.
\end{itemize}

\subsection{Indoor Hotspot Scenario}
Measurements were performed in the open office environment of the NYU WIRELESS Research Center. Total of 20 TX-RX locations were measured covering the entire office space spanning distances between 11 and 97 m. Cubicles, offices, labs, and conference rooms in the research center were partitioned with drywall and glass panels with wooden or glass doors.  A map of the environment is shown in Fig. \ref{fig:Map_0PAS} (a) \cite{Shakya2024ojcoms,Shakya2024gc}.

\subsection{Measurement Procedure}
The channel sounder calibration process ensures accurate capture of the multipath power, delay, and direction, and is performed at the start and end of each day of propagation measurements. At each TX-RX location pair, the strongest azimuth and elevation pointing directions are determined by carefully observing changes in the recorded power level as the TX and RX antennas are moved in one-degree steps. The careful stepping ensures capturing the strongest MPC with maximum power. Keeping the same zenith of departure (ZOD) elevation angle, TX angles of departure (AOD) with significant RX received power are determined through rapid scans, as detailed in \cite{Shakya2024ojcoms}. 

For each TX AOD pointing direction, the RX is swept 360$^{\circ}$ in the azimuth in antenna HPBW steps. Following the stepped sweep at boresight 
 zenith of arrival (ZOA) elevation, the RX is up and down tilted by the antenna HPBW and swept again across the azimuth. Next, the ZOD elevation is changed by down-tilting the TX by the antenna HPBW. RX azimuth sweeps in HPBW steps are performed for the down-tilted TX ZOD at RX boresight and HPBW down-tilted ZOAs. 

\section{Angular Statistics meausured at 6.75 and 16.95 GHz}
\label{sxn}


The double directional channel impulse response for multipath propagation between the TX and RX can be defined with \eqref{eq_omni} \cite{Steinbauer2001apm}.
\begin{equation}
\label{eq_omni}
\begin{aligned}
h_{omni}(t,\overrightarrow{\Theta},\overrightarrow{\Phi})= &  \sum_{n=1}^{N}\sum_{m=1}^{M_n}a_{m,n}e^{j\varphi_{m,n}}\cdot \delta(t-\tau_{m,n}) \\& \cdot \delta(\overrightarrow{\Theta}-\overrightarrow{\Theta_{m,n}})\cdot\delta(\overrightarrow{\Phi}-\overrightarrow{\Phi_{m,n}}),
\end{aligned}
\end{equation}
where $t$ is the absolute propagation time, $\overrightarrow{\Theta}=\left(\phi_{AOD},\theta_{ZOD}\right)$ represents the 3D TX pointing direction vector, and $\overrightarrow{\Phi}=\left(\phi_{AOA},\theta_{ZOA}\right)$ is the RX pointing direction vector. $N$ and $M_n$ denote the number of time clusters (TCs) and the number of cluster subpaths (SPs), respectively, as defined in \cite{Shakya2024ojcoms}; $a_{m,n}$ is the magnitude of the $m^{th}$ SP belonging to the $n^{th}$ TC, while $\varphi_{m,n}$ and $\tau_{m,n}$ represent the phase and propagation delay of the SP, respectively. Likewise, $\overrightarrow{\Theta_{m,n}}$ and $\overrightarrow{\Phi_{m,n}}$ are the vectors representing AOD/ZOD and azimuth of arrival (AOA)/ZOA for the SP, respectively. The terms MPC and SP are used interchangeably \cite{Samimi2016tmtt}.

The propagation behavior of the multipath is characterized through primary statistics including the number of time clusters and spatial lobes, cluster delay and multipath delay, received power of each SP in a cluster, and direction of arrival and departure of SPs. The secondary statistics for temporal and spatial propagation behavior including the RMS angular spread (AS) at both TX and RX help characterize the wireless channel at a large scale and are crucial parameters for cellular network and radio system designs. The RMS AS provides a measure of the spatial dispersion of angles at which MPCs arrive or depart from a receiver or transmitter. The azimuth angular spread of arrival (ASA) and the zenith angular spread of arrival (ZSA) characterize the spread of multipath arriving at the RX in horizontal and vertical planes, respectively. Similarly, at the TX side, the azimuth angular spread of departure (ASD) and the zenith angular spread of departure (ZSD) describe the spread of MPCs being transmitted over the wireless channel that are captured at the RX. 

The NYU WIRELESS channel measurements include directional measurements, which are carefully combined to synthesize the omnidirectional antenna pattern, removing antenna gains \cite{sun2015gc,Shakya2024ojcoms,Shakya2024gc}. 
The omni directional antenna patterns facilitate the implementation of arbitrary antenna patterns for simulations, with 3GPP providing omnidirectional models \cite{3GPPTR38901,rappaport2017overview}. 

\begin{figure}[htbp]
    \centering
    \begin{subfigure}[t]{0.98\linewidth}
        \centering
        \includegraphics[width=\textwidth]{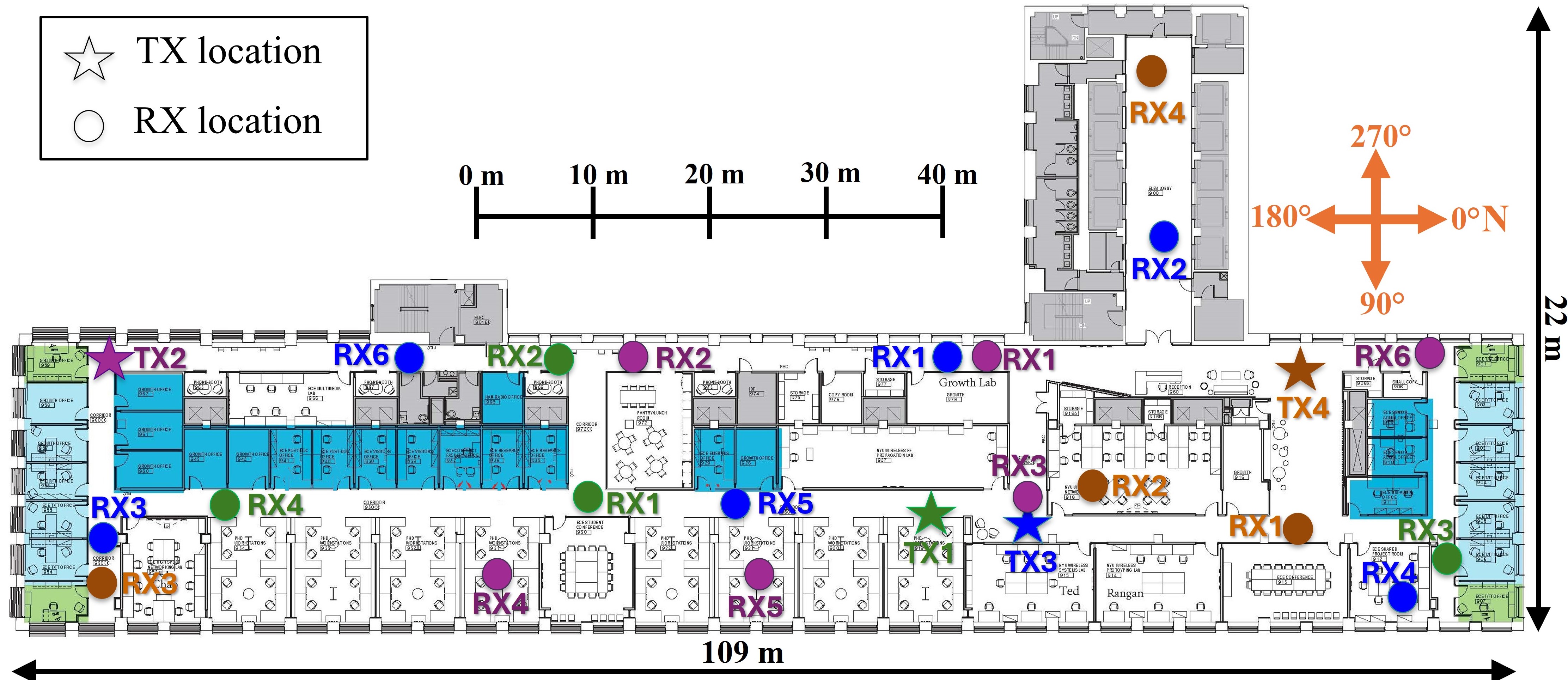}
        \caption{Map of the indoor environment.}
        \label{fig:Map}
    \end{subfigure}\\
    \vspace{5 pt}
    \begin{subfigure}[t]{0.6\linewidth}
        \centering
        \includegraphics[width=\textwidth]{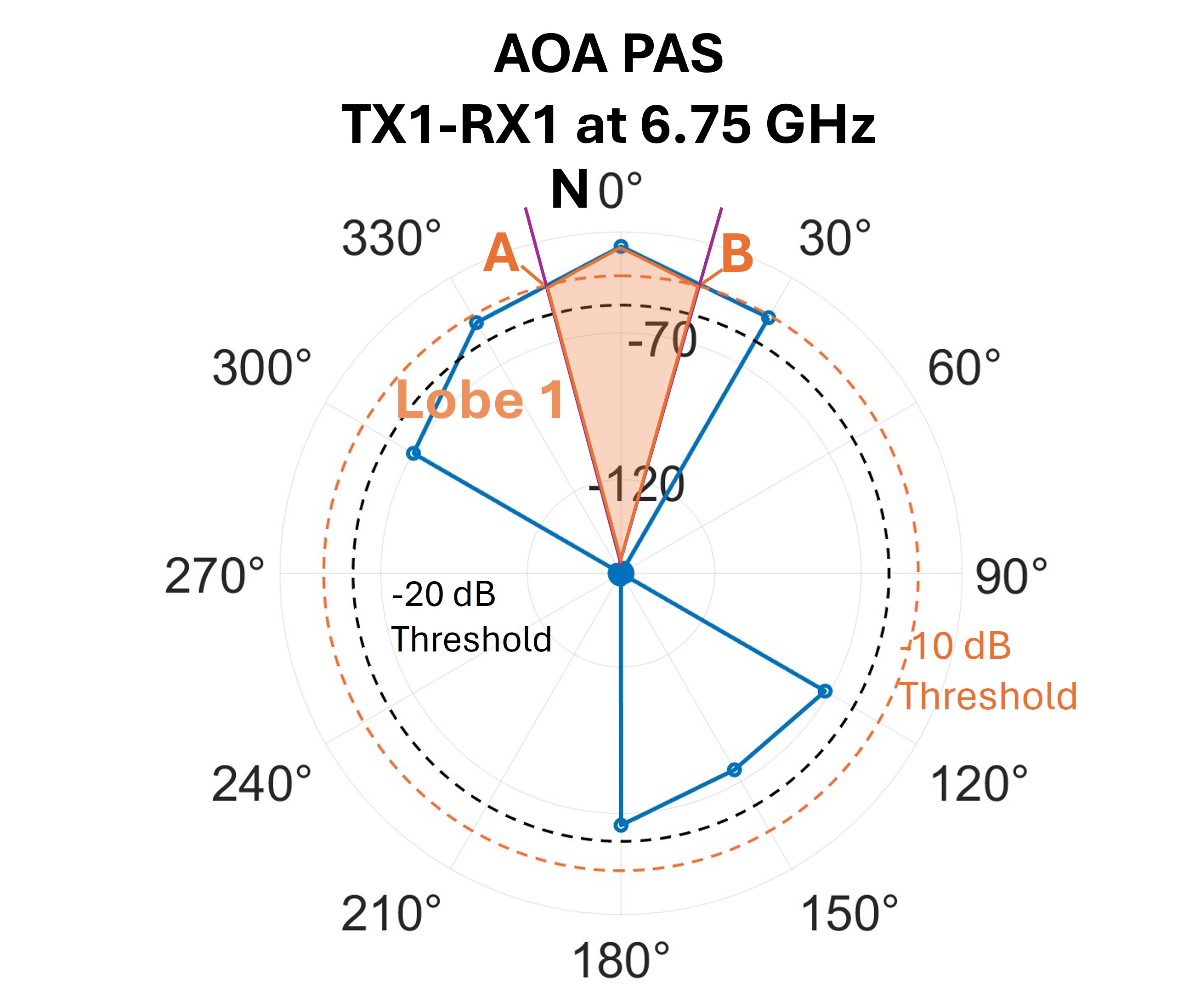}
        \caption{AOA PAS at 6.75 GHz.}
    \end{subfigure}
    
    \caption{PAS illustration highlighting a spatial lobe.}
    \label{fig:Map_0PAS}
    \vspace{-10pt}
\end{figure}

\textcolor{black}{The blue dots on the PAS in Fig. \ref{fig:Map_0PAS} (b) represent the channel sounder RX AOA pointing directions during the directional measurements at TX1-RX1 location at 6.75 GHz. The AOA/AOD powers between the measured directions are obtained through linear interpolation of the measured powers in linear scale, as the PAS has a spatial resolution corresponding to the antenna HPBW. Considering the PAS in Fig. \ref{fig:Map_0PAS} (b), a spatial lobe threshold (SLT) is defined at 10 dB below the peak power in the PAS (``-10 dB Threshold" orange dashed line in Fig. \ref{fig:Map_0PAS} (b)) for identifying spatial lobes (SL).}

\begin{figure}[htbp]
    \centering
    \begin{subfigure}[t]{0.45\linewidth}
        \centering
        \includegraphics[width=\textwidth]{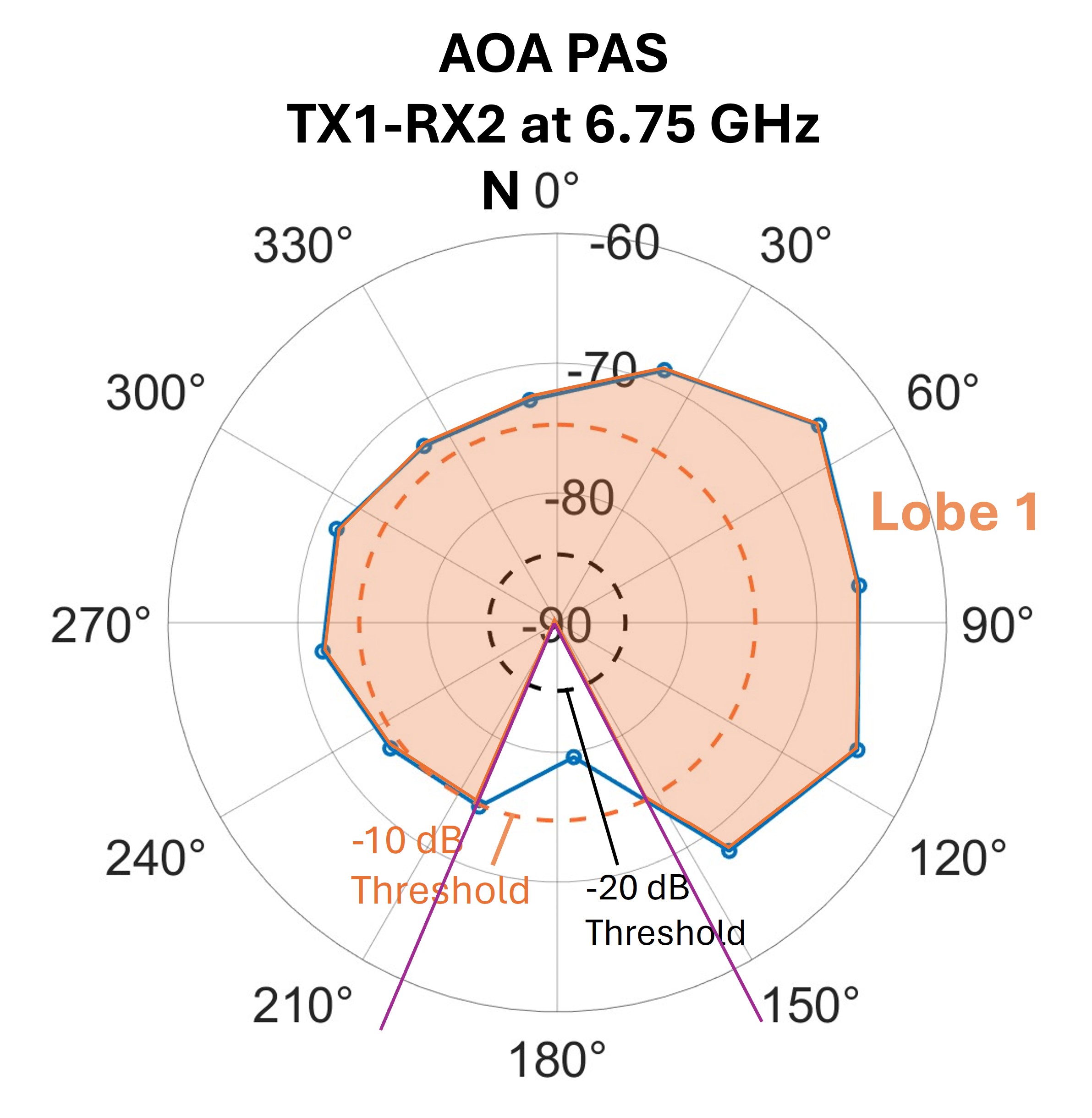}
        \caption{AOA PAS at 6.75 GHz.}
    \end{subfigure}
    \begin{subfigure}[t]{0.47\linewidth}
        \centering
        \includegraphics[width=\textwidth]{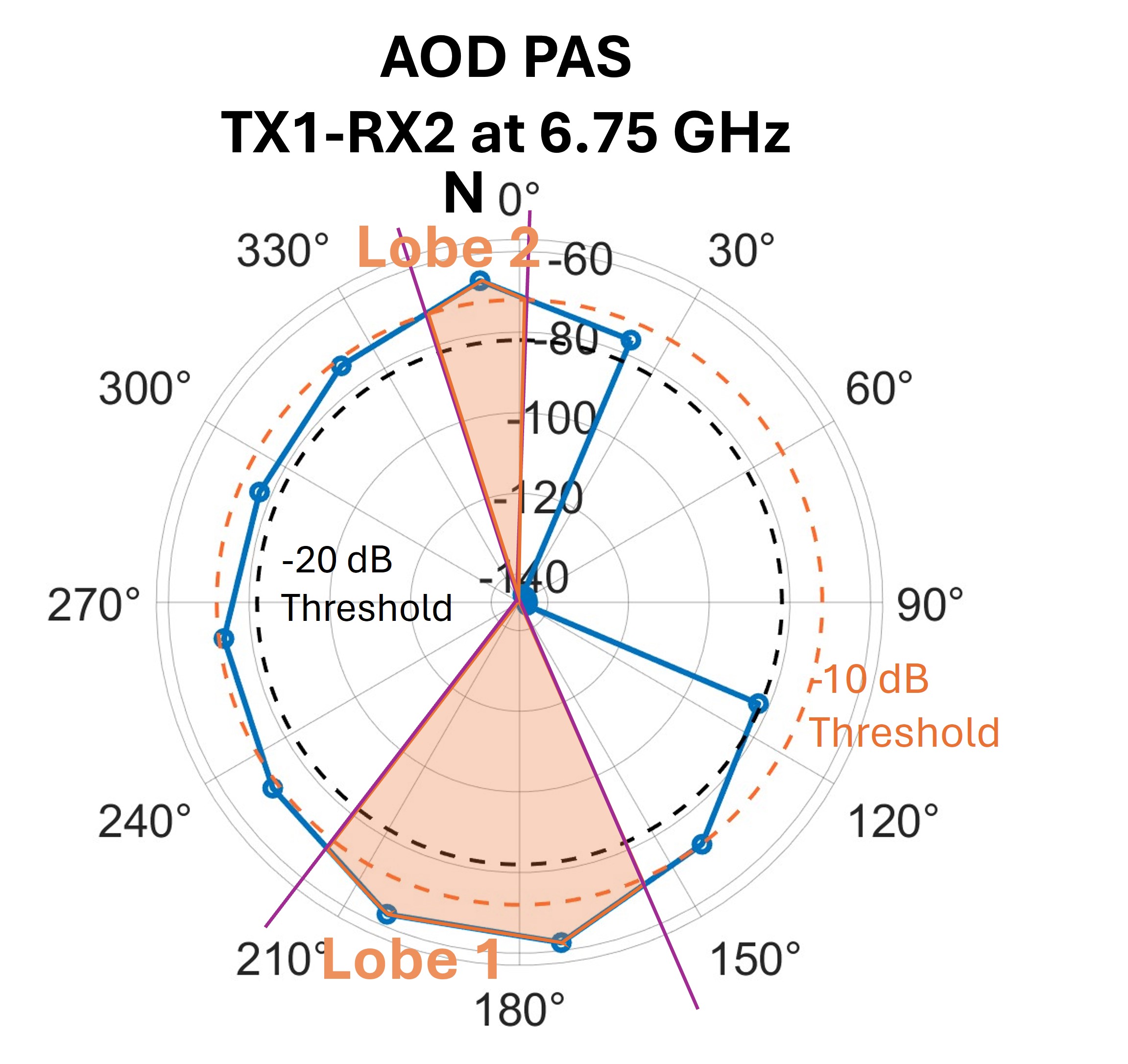}
        \caption{AOD PAS at 6.75 GHz.}
    \end{subfigure}\\
    
    \begin{subfigure}[t]{0.47\linewidth}
        \centering
        \includegraphics[width=\textwidth]{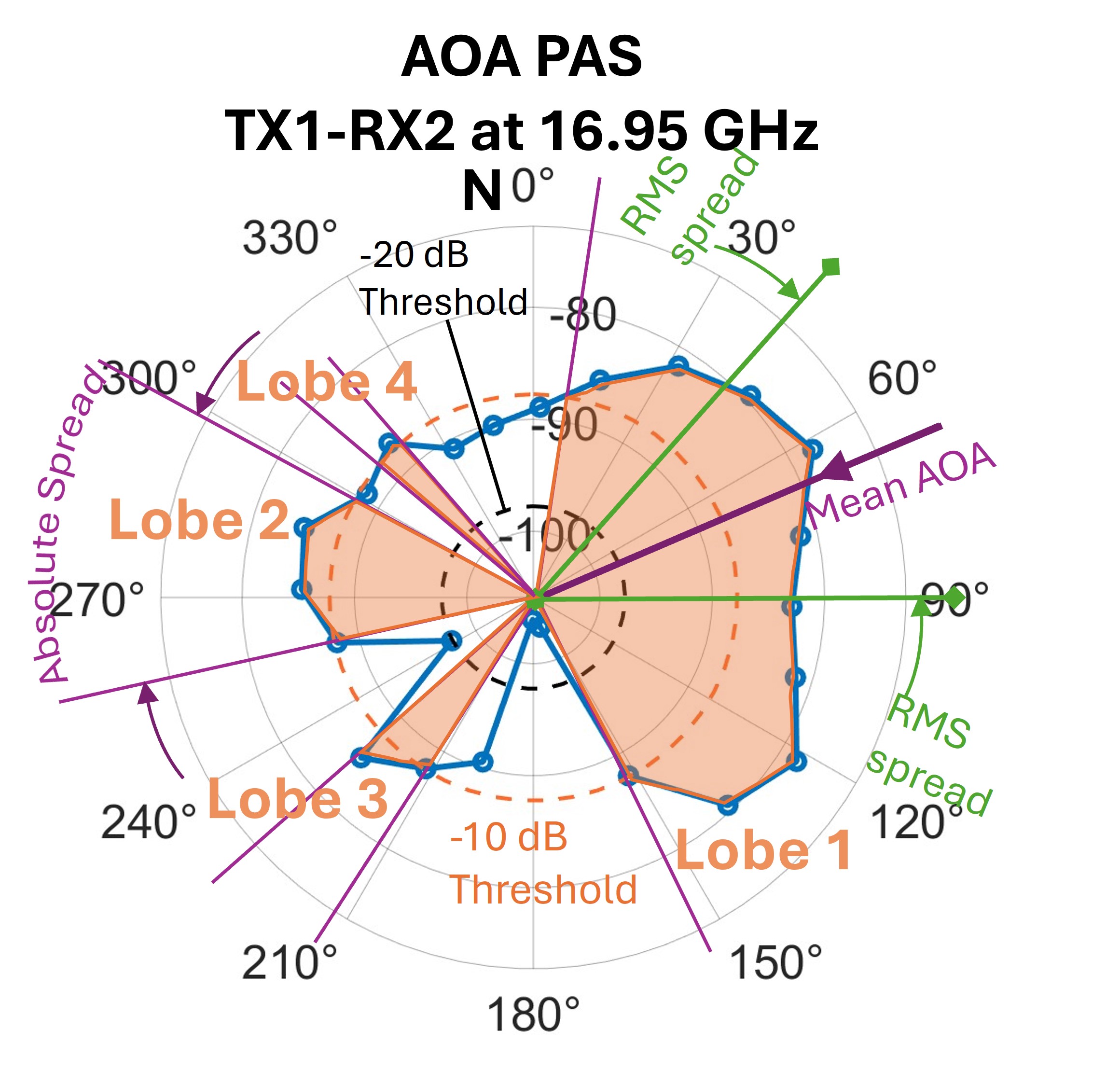}
        \caption{AOA PAS at 16.95 GHz.}
    \end{subfigure}
    \begin{subfigure}[t]{0.47\linewidth}
        \centering
        \includegraphics[width=\textwidth]{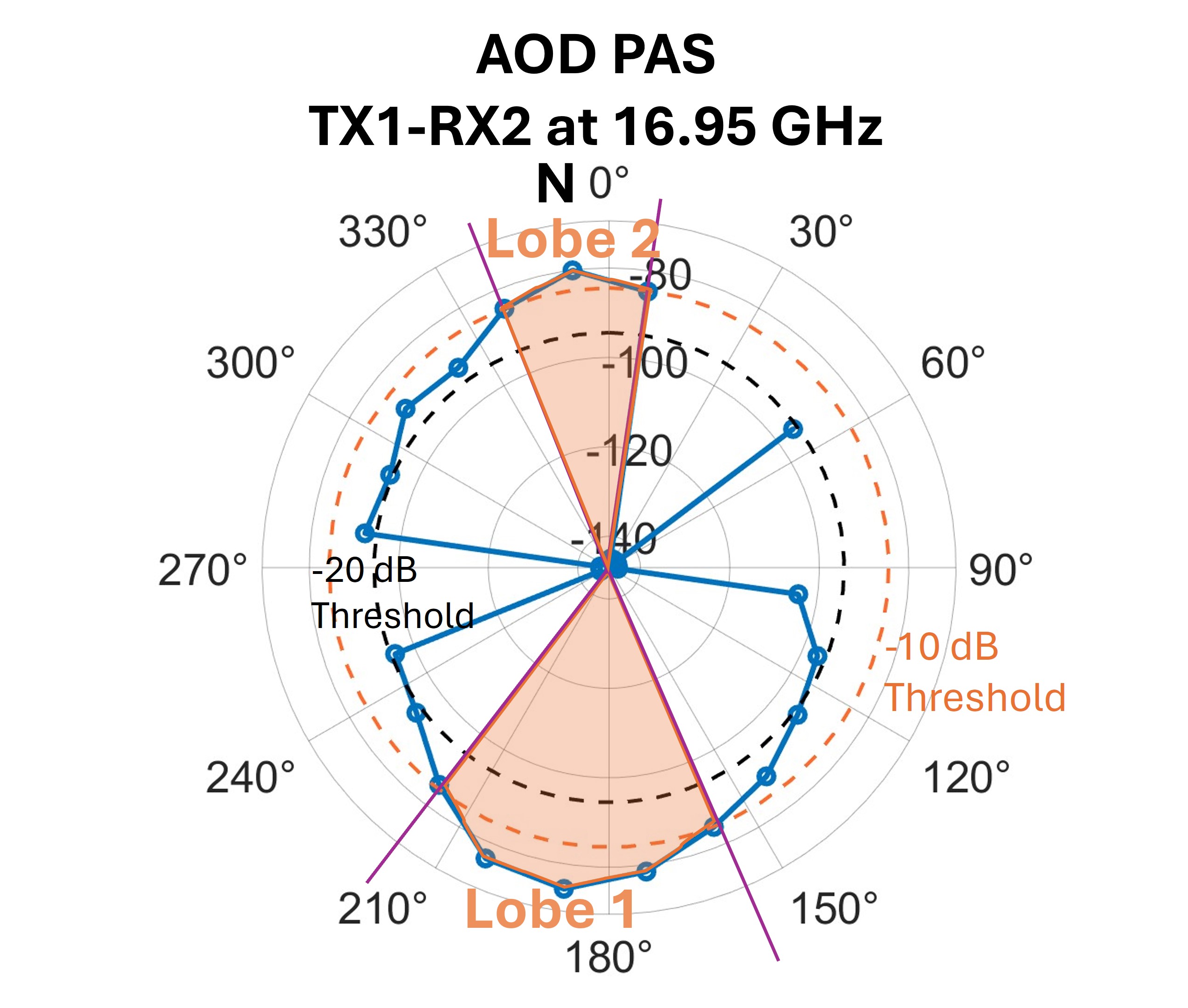}
        \caption{AOD PAS at 16.95 GHz.}
    \end{subfigure}
    
    \caption{PAS illustration highlighting spatial lobes.}
    \vspace{-10 pt}
    \label{fig:17_PAS}
\end{figure}

\textcolor{black}{Using the same ``-10 dB Threshold" on the PAS as the SLT, a spatial lobe is defined as contiguous spatial directions with powers above the SLT. An example is illustrated in Fig. \ref{fig:Map_0PAS} (b), where the SL begins at angle `A' and spans all directions up to angle `B'. The evaluation of the RMS AS considers the multipath departure/arrival directions and powers corresponding to angles `A' and `B', and all antenna pointing directions within the SLs represented by the blue dots. The SLs in a PAS are illustrated in Fig. \ref{fig:Map_0PAS} (b) ---a narrow SL encompassing only one antenna pointing direction at 0 $^\circ$--- and Fig. \ref{fig:17_PAS} ---diverse and relatively wider SLs comprised of multiple pointing directions--- as the orange-filled region with received powers above the SLT. As an example, Fig. \ref{fig:17_PAS} (a) and (b) show the AOA and AOD PAS at 6.75 GHz, while Fig. \ref{fig:17_PAS} (c) and (d) show AOA and AOD PAS at 16.95 GHz for the same TX1-RX2 location indicated in Fig. \ref{fig:Map_0PAS} (a). The shape of the PASs are visibly similar but have different numbers of lobes between the two frequencies. 6.75 GHz TX1-RX2 AOA PAS in Fig. \ref{fig:17_PAS} (a) is a single lobe with more power focused around 110$^{\circ}$, while 16.95 GHz PAS in Fig. \ref{fig:17_PAS} (c) has four spatial lobes.} 

\textcolor{black}{Based on circular statistics, the omnidirectional RMS AS is evaluated in radians using \eqref{eq_asOmni} for the circular standard deviation of the MPC departure/arrival angles considering all SLs in the PAS\cite{3GPPTR38901}. AS for each location is presented in \cite{Ted2025icc}}
\begin{equation}
\label{eq_asOmni}
\begin{aligned}
AS_{omni} = \sqrt{-2\times ln{\left|\frac{\sum_{l=1}^{L}\sum_{m=1}^{M_l}e^{(j(\phi_{l,m}\text{ or }\theta_{l,m}))} a^2_{l,m}}{\sum_{l=1}^{L}\sum_{m=1}^{M_l} a^2_{l,m}}\right|}},
\end{aligned}
\end{equation}
$\phi_{l,m}$ in \eqref{eq_asOmni} can represent the AOA or AOD and $\theta_{l,m}$ can represent ZOA or ZOD for the $m^{th}$ MPC in the $l^{th}$ SL. Evaluating the AS for each SL defined from the PAS, as shown in Fig. \ref{fig:Map_0PAS}, results in the lobe AS for the SL. For each SL, the lobe AS is evaluated using \eqref{eq_asLobe}, as the circular standard deviation of the departure/arrival directions of MPCs within the SL\cite{3GPPTR38901}.   

\begin{equation}
\label{eq_asLobe}
\begin{aligned}
AS_{lobe,l} = \sqrt{-2\times ln{\left|\frac{\sum_{m=1}^{M_l}e^{(j(\phi_{m}\text{ or }\theta_{m}))} a^2_{m}}{\sum_{m=1}^{M_l} a^2_{m}}\right|}},
\end{aligned}
\end{equation}

\section{Angular Statistics of 3GPP TR38.901 Models}
\label{3GPP}
\textcolor{black}{Table \ref{tab:AS} compares the RMS AS evaluated from the InH measurements, with the results from 3GPP TR 38.901 models for 0.5-100 GHz \cite{3GPPTR38901}. The mean ($\mu$) and standard deviation ($\sigma$) values are calculated for the $\log_{10}$ logarithm of the observed AS values at the T-R locations for a direct comparison to the 3GPP omnidirectional RMS AS considering a log-normal distribution of the RMS AS. }

\begin{equation}
\label{eq_meanAS}
\begin{aligned}
\color{black}
\mu_{\lg \textit{Lobe AS}} = \frac{\sum_l \log_{10}(\text{Lobe AS}_l)}{L},\\
\color{black}
\mu_{\lg \textit{Omni AS}} = \frac{\sum_n \log_{10}(\text{Omni AS}_n)}{N}
\end{aligned}
\end{equation}
 where, $L$ denotes the total SLs and $N$ denotes the total number of TR locations.


 \textcolor{black}{The evaluated AS values are also expressed in Table \ref{tab:AS} as the expectation of the log-normal distribution ($\mathbb{E}(\mathbf{X})=10^{\mu+\frac{\sigma^2}{2}}$) in degrees ($^\circ$) for both the NYU measurements and 3GPP results for a physically intuitive comparison.}

3GPP calculates the omnidirectional RMS AS using a formulation identical to \eqref{eq_asOmni}. Based on the evaluation of AS across various frequencies, 3GPP provides frequency-dependent models to characterize AS up to 100 GHz. These models are specifically detailed in \cite{3GPPTR38901} under ``Table 7.5-6 Part-2" and ``Table 7.5-10."

\setlength{\tabcolsep}{10pt} 
\begin{table*}[!t]
    \centering
    \scriptsize
    \caption{Angular Spread Characteristics measured by NYU WIRELESS at 6.75 GHz and 16.95 GHz in LOS and NLOS for the InH environment \cite{Shakya2024ojcoms,Shakya2024gc} and Comparison to 3GPP Models \cite{3GPPTR38901}.}
    \label{tab:AS}
    \renewcommand{\arraystretch}{1.29} 
    \begin{tabular}{|@{\hspace{5 pt}}P@{\hspace{15 pt}}|e|c|z|x|x|e?c|z|x|x|e|}
        \specialrule{1.2 pt}{0 pt}{0 pt}

        \multirow{2}{*}{\textbf{Metric}} & \multirow{2}{*}{\textbf{Condition}} & \multicolumn{5}{c?}{\textbf{6.75 GHz}} & \multicolumn{5}{c|}{\textbf{16.95 GHz}} \\
        \cline{3-12}
        & & \textbf{NYU} & \textbf{3GPP} & $\mathbb{E}(\textbf{NYU})$ & $\mathbb{E}(\textbf{3GPP})$  & \textbf{$|\Delta_{\text{\tiny NYU-3GPP}}|$} & \textbf{NYU} & \textbf{3GPP} & $\mathbb{E}(\textbf{NYU})$ & $\mathbb{E}(\textbf{3GPP})$  & \textbf{$|\Delta_{\text{\tiny NYU-3GPP}}|$} \\
        \specialrule{1 pt}{0 pt}{0 pt}

        \multirow{4}{*}{\shortstack{Lobe RMS $\lg_\text{ASA}$\\ = $\log_{10}$($\text{ASA}/1^{\circ}$)}}
        & $\mu_\text{lgASA}^\text{LOS}$ & 1.02 & -- & \multirow{2}{*}{10.75$^\circ$} & \multirow{2}{*}{--} & \multirow{2}{*}{--}  & 0.85 & -- & \multirow{2}{*}{7.10$^\circ$} & \multirow{2}{*}{--}  & \multirow{2}{*}{--}  \\ 
        \cline{2-3} \cline{4-4} \cline{8-8} \cline{9-9}  
        & $\sigma_\text{lgASA}^\text{LOS}$ & 0.15 & -- &  &  &  & 0.05 & -- &  &  &  \\ \cline{2-12}
        & $\mu_\text{lgASA}^\text{NLOS}$ & 1.42 & -- & \multirow{2}{*}{28.98$^\circ$}  & \multirow{2}{*}{--}  & \multirow{2}{*}{--}  & 1.05 & -- & \multirow{2}{*}{11.86$^\circ$}  & \multirow{2}{*}{--}  & \multirow{2}{*}{--}  \\ \cline{2-3} \cline{4-4} \cline{8-8} \cline{9-9}  
        & $\sigma_\text{lgASA}^\text{NLOS}$ & 0.29 & -- &  &  &  & 0.22 & -- &  &  &  \\
        \specialrule{1 pt}{0 pt}{0 pt}
        
        \multirow{4}{*}{\shortstack{Omni RMS $\lg_\text{ASA}$\\  $=\log_{10}$($\text{ASA}/1^{\circ}$)}}
        & $\mu_\text{lgASA}^\text{LOS}$& 1.54 & 1.61 & \multirow{2}{*}{41.31$^\circ$} & \multirow{2}{*}{43.07$^\circ$} & \multirow{2}{*}{1.76$^\circ$} & 1.14 & 1.54 & \multirow{2}{*}{17.25$^\circ$} & \multirow{2}{*}{37.48$^\circ$} & \multirow{2}{*}{20.23$^\circ$} \\ \cline{2-3} \cline{4-4} \cline{8-8} \cline{9-9} 
        & $\sigma_\text{lgASA}^\text{LOS}$ & 0.39 & 0.22 &  &  &  & 0.44 & 0.26 &  &  &  \\ \cline{2-12}
        & $\mu_\text{lgASA}^\text{NLOS}$ & 1.74 & 1.77 & \multirow{2}{*}{58.10$^\circ$} & \multirow{2}{*}{60.88$^\circ$} &\multirow{2}{*}{2.77$^\circ$} & 1.68 & 1.73 & \multirow{2}{*}{54.26$^\circ$} & \multirow{2}{*}{56.50$^\circ$} & \multirow{2}{*}{2.24$^\circ$} \\ \cline{2-3} \cline{4-4} \cline{8-8} \cline{9-9} 
        & $\sigma_\text{lgASA}^\text{NLOS}$ & 0.22 & 0.17 &  &  &  & 0.33 & 0.21 &  &  &  \\ \specialrule{1 pt}{0 pt}{0 pt}
        
        \multirow{4}{*}{\shortstack{Lobe RMS $\lg_\text{ASD}$\\=$\log_{10}$($\text{ASD}/1^{\circ}$)}}
        & $\mu_\text{lgASD}^\text{LOS}$ & 0.98 & -- & \multirow{2}{*}{9.83$^\circ$} & \multirow{2}{*}{--} & \multirow{2}{*}{--} & 0.78 & -- & \multirow{2}{*}{6.21$^\circ$} & \multirow{2}{*}{--} & \multirow{2}{*}{--} \\ \cline{2-3} \cline{4-4} \cline{8-8} \cline{9-9} 
        & $\sigma_\text{lgASD}^\text{LOS}$ & 0.21 & -- &  &  &  & 0.16 & -- &  &  &  \\ \cline{2-12}
        & $\mu_\text{lgASD}^\text{NLOS}$ & 1.08 & -- & \multirow{2}{*}{12.71$^\circ$} & \multirow{2}{*}{--} & \multirow{2}{*}{--} & 1.02 & -- & \multirow{2}{*}{10.75$^\circ$} & \multirow{2}{*}{--} & \multirow{2}{*}{--} \\ \cline{2-3} \cline{4-4} \cline{8-8} \cline{9-9} 
        & $\sigma_\text{lgASD}^\text{NLOS}$ & 0.22 & -- &  &  &  & 0.15 & -- &  &  &  \\ \specialrule{1 pt}{0 pt}{0 pt}
        
        \multirow{4}{*}{\shortstack{Omni RMS $\lg_\text{ASD}$\\=$\log_{10}$($\text{ASD}/1^{\circ}$)}}
        & $\mu_\text{lgASD}^\text{LOS}$ & 1.84 & 1.60 & \multirow{2}{*}{70.34$^\circ$} & \multirow{2}{*}{41.32$^\circ$} & \multirow{2}{*}{29.02$^\circ$} & 1.79 & 1.60 & \multirow{2}{*}{62.52$^\circ$} & \multirow{2}{*}{41.32$^\circ$} & \multirow{2}{*}{21.20$^\circ$} \\ \cline{2-3} \cline{4-4} \cline{8-8} \cline{9-9} 
        & $\sigma_\text{lgASD}^\text{LOS}$ & 0.12 & 0.18 &  &  &  & 0.11 & 0.18 &  &  &  \\ \cline{2-12}
        & $\mu_\text{lgASD}^{\text{NLOS}}$ & 1.67 & 1.62 & \multirow{2}{*}{48.76$^\circ$} & \multirow{2}{*}{44.80$^\circ$} & \multirow{2}{*}{3.96$^\circ$} & 1.68 & 1.62 & \multirow{2}{*}{53.85$^\circ$} & \multirow{2}{*}{44.80$^\circ$} & \multirow{2}{*}{9.05$^\circ$} \\ \cline{2-3} \cline{4-4} \cline{8-8} \cline{9-9} 
        & $\sigma_\text{lgASD}^\text{NLOS}$ & 0.19 & 0.25 &  &  &  & 0.32 & 0.25 &  &  &  \\ \specialrule{1 pt}{0 pt}{0 pt}
        
        \multirow{4}{*}{\shortstack{Lobe RMS $\lg_\text{ZSA}$\\=$\log_{10}$($\text{ZSA}/1^{\circ}$)}}
        & $\mu_{\text{lgZSA}}^\text{LOS}$ & 0.72 & -- & \multirow{2}{*}{5.67$^\circ$} &  \multirow{2}{*}{--}  & \multirow{2}{*}{--}  & 0.77 & -- & \multirow{2}{*}{6.00$^\circ$} &  \multirow{2}{*}{--}  &  \multirow{2}{*}{--}  \\ \cline{2-3} \cline{4-4} \cline{8-8} \cline{9-9} 
        & $\sigma_{\text{lgZSA}}^\text{LOS}$ & 0.26 & -- &  &  &  & 0.13 & -- &  &  &  \\ \cline{2-12}
        & $\mu_\text{lgZSA}^\text{NLOS}$ & 0.64 & -- & \multirow{2}{*}{4.84$^\circ$} &  \multirow{2}{*}{--}  &  \multirow{2}{*}{--} & 0.68 & -- & \multirow{2}{*}{5.11$^\circ$} &  \multirow{2}{*}{--}  &  \multirow{2}{*}{--}  \\ \cline{2-3} \cline{4-4} \cline{8-8} \cline{9-9} 
        & $\sigma_\text{lgZSA}^\text{NLOS}$  & 0.30 & -- &  &  &  & 0.24 & -- &  &  &  \\ \specialrule{1 pt}{0 pt}{0 pt}
        
        \multirow{4}{*}{\shortstack{Omni RMS $\lg_\text{ZSA}$\\=$\log_{10}$($\text{ZSA}/1^{\circ}$)}}
        & $\mu_\text{lgZSA}^\text{LOS}$ & 1.21 & 1.21 & \multirow{2}{*}{16.31$^\circ$} & \multirow{2}{*}{17.20$^\circ$} & \multirow{2}{*}{0.89$^\circ$} & 1.01 & 1.11 & \multirow{2}{*}{10.26$^\circ$} & \multirow{2}{*}{13.58$^\circ$} & \multirow{2}{*}{3.32$^\circ$} \\ \cline{2-3} \cline{4-4} \cline{8-8} \cline{9-9} 
        & $\sigma_\text{lgZSA}^\text{LOS}$ & 0.07 & 0.23 &  &  &  & 0.05 & 0.21 &  &  &  \\ \cline{2-12}
        & $\mu_\text{lgZSA}^\text{NLOS}$ & 1.01 & 1.25 & \multirow{2}{*}{11.27$^\circ$} & \multirow{2}{*}{29.82$^\circ$} & \multirow{2}{*}{18.54$^\circ$} & 0.96 & 1.20 & \multirow{2}{*}{9.23$^\circ$} & \multirow{2}{*}{25.03$^\circ$} & \multirow{2}{*}{15.8$^\circ$} \\ \cline{2-3} \cline{4-4} \cline{8-8} \cline{9-9} 
        & $\sigma_\text{lgZSA}^\text{NLOS}$ & 0.29 & 0.67 &  &  &  & 0.10 & 0.63 &  &  &  \\ \specialrule{1 pt}{0 pt}{0 pt}
        
        \multirow{4}{*}{\shortstack{Lobe RMS $\lg_\text{ZSD}$\\=$\log_{10}$($\text{ZSD}/1^{\circ}$)}}
        & $\mu_\text{lgZSD}^\text{LOS}$ & 0.64 & -- & \multirow{2}{*}{4.50$^\circ$} &  \multirow{2}{*}{--}  &  \multirow{2}{*}{--}  & 0.68 & -- & \multirow{2}{*}{4.82$^\circ$} &  \multirow{2}{*}{--}  &  \multirow{2}{*}{--}  \\ \cline{2-3} \cline{4-4} \cline{8-8} \cline{9-9} 
        & $\sigma_\text{lgZSD}^\text{LOS}$ & 0.16 & -- &  &  &  & 0.08 & -- &  &  &  \\ \cline{2-12}
        & $\mu_\text{lgZSD}^\text{NLOS}$  & 0.74 & -- &  \multirow{2}{*}{5.73$^\circ$}  &  \multirow{2}{*}{--}  &  \multirow{2}{*}{--}  & 0.72 & -- &  \multirow{2}{*}{5.45$^\circ$}  &  \multirow{2}{*}{--}  &  \multirow{2}{*}{--}  \\ \cline{2-3} \cline{4-4} \cline{8-8} \cline{9-9} 
        & $\sigma_\text{lgZSD}^\text{NLOS}$  & 0.19 & -- &  &  &  & 0.18 & -- &  &  &  \\ \specialrule{1 pt}{0 pt}{0 pt}

        \multirow{4}{*}{\shortstack{Omni RMS $\lg_\text{ZSD}$\\=$\log_{10}$($\text{ZSD}/1^{\circ}$)}}
        & $\mu_\text{lgZSD}^\text{LOS}$  & 1.07 & 0.96 & \multirow{2}{*}{11.80$^\circ$} & \multirow{2}{*}{11.17$^\circ$} &  \multirow{2}{*}{0.62$^\circ$}  & 0.91 & 0.43 & \multirow{2}{*}{8.15$^\circ$} & \multirow{2}{*}{3.43$^\circ$} &  \multirow{2}{*}{4.72$^\circ$}  \\ \cline{2-3} \cline{4-4} \cline{8-8} \cline{9-9} 
        & $\sigma_\text{lgZSD}^\text{LOS}$ & 0.06 & 0.42 &  &  &  & 0.05 & 0.46 &  &  &  \\ \cline{2-12}
        & $\mu_\text{lgZSD}^\text{NLOS}$  & 1.07 & 1.08 &  \multirow{2}{*}{11.91$^\circ$}  &  \multirow{2}{*}{13.96$^\circ$}  &  \multirow{2}{*}{2.04$^\circ$}  & 0.89 & 1.08 &  \multirow{2}{*}{7.94$^\circ$} &  \multirow{2}{*}{13.96$^\circ$}  &  \multirow{2}{*}{6.02$^\circ$}  \\ \cline{2-3} \cline{4-4} \cline{8-8} \cline{9-9} 
        & $\sigma_\text{lgZSD}^\text{NLOS}$  & 0.11 & 0.36 &  &  &  & 0.14 & 0.36 &  &  &  \\ \specialrule{1.2 pt}{0 pt}{0 pt}

    \end{tabular}
    \vspace{-15pt}
\end{table*}
\renewcommand{\arraystretch}{1}

\begin{figure}[t]
    \centering
    \begin{subfigure}[t]{0.7\linewidth}
        \centering
        \includegraphics[width=\textwidth]{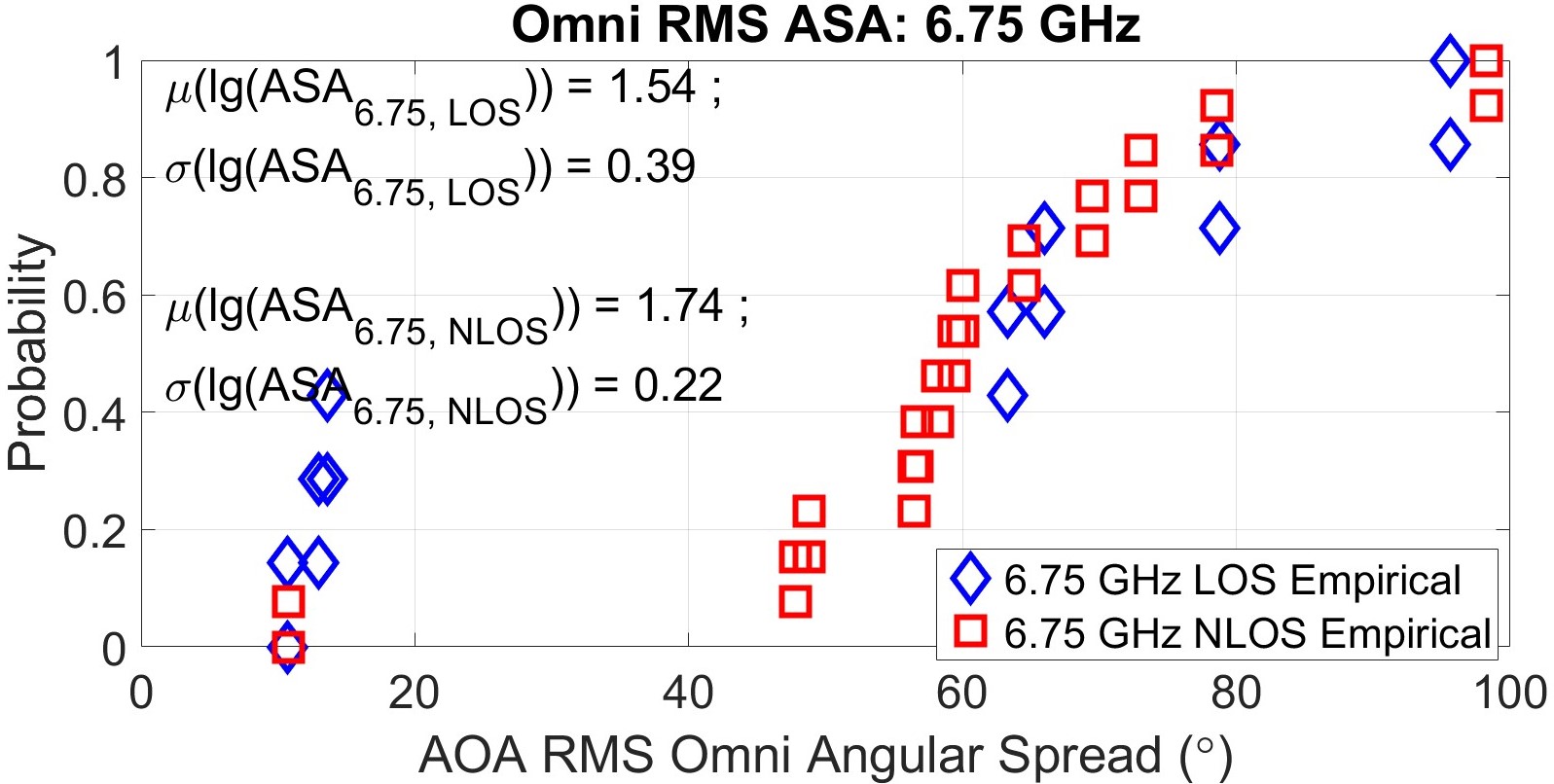}
        \caption{Omni ASA at 6.75 GHz.}
    \end{subfigure}\\
    \vspace{5pt}
    \begin{subfigure}[t]{0.7\linewidth}
        \centering
        \includegraphics[width=\textwidth]{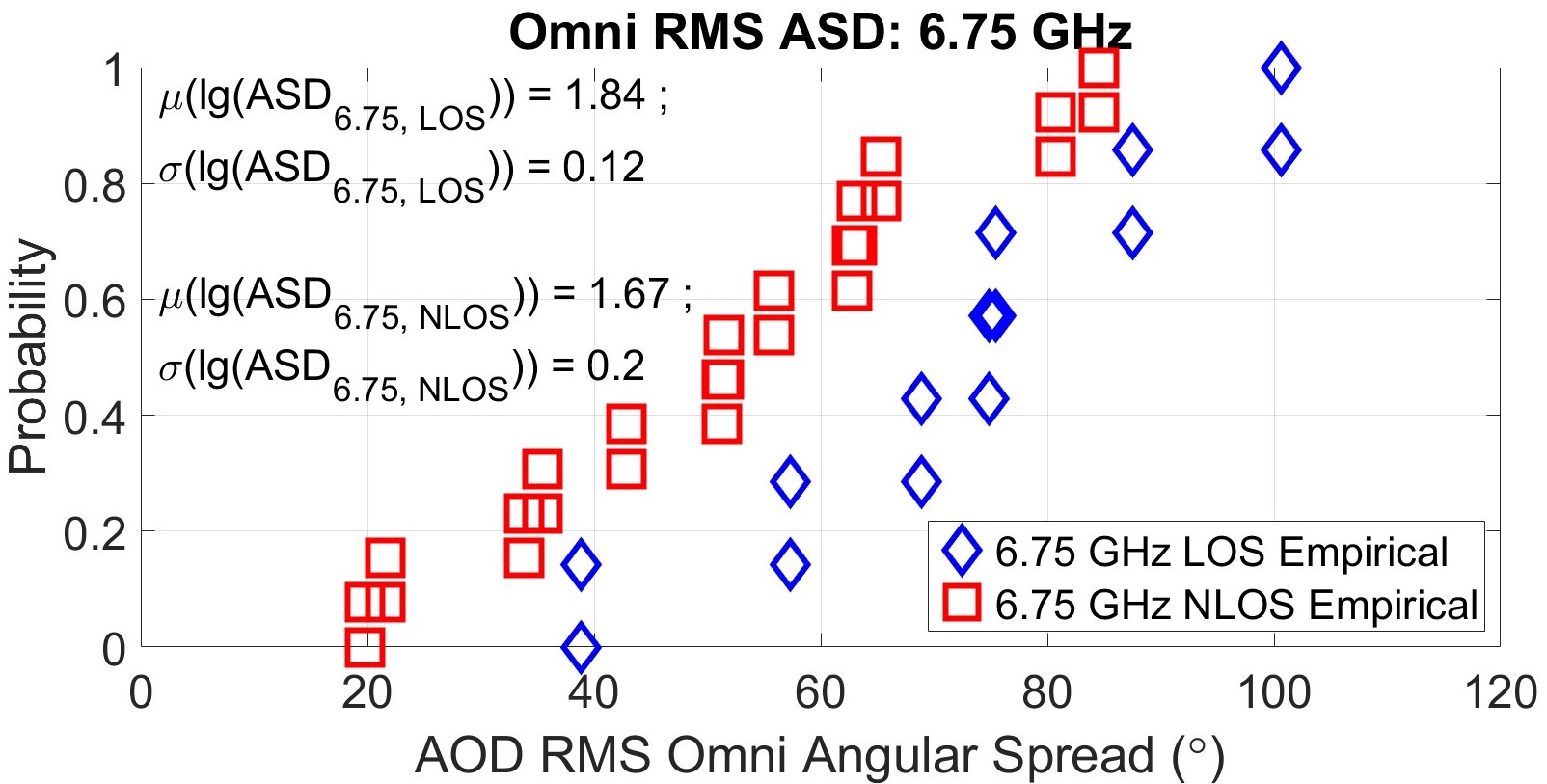}
        \caption{Omni ASD at 6.75 GHz.}
    \end{subfigure}\\
    \vspace{5pt}
    \begin{subfigure}[t]{0.7\linewidth}
        \centering
        \includegraphics[width=\textwidth]{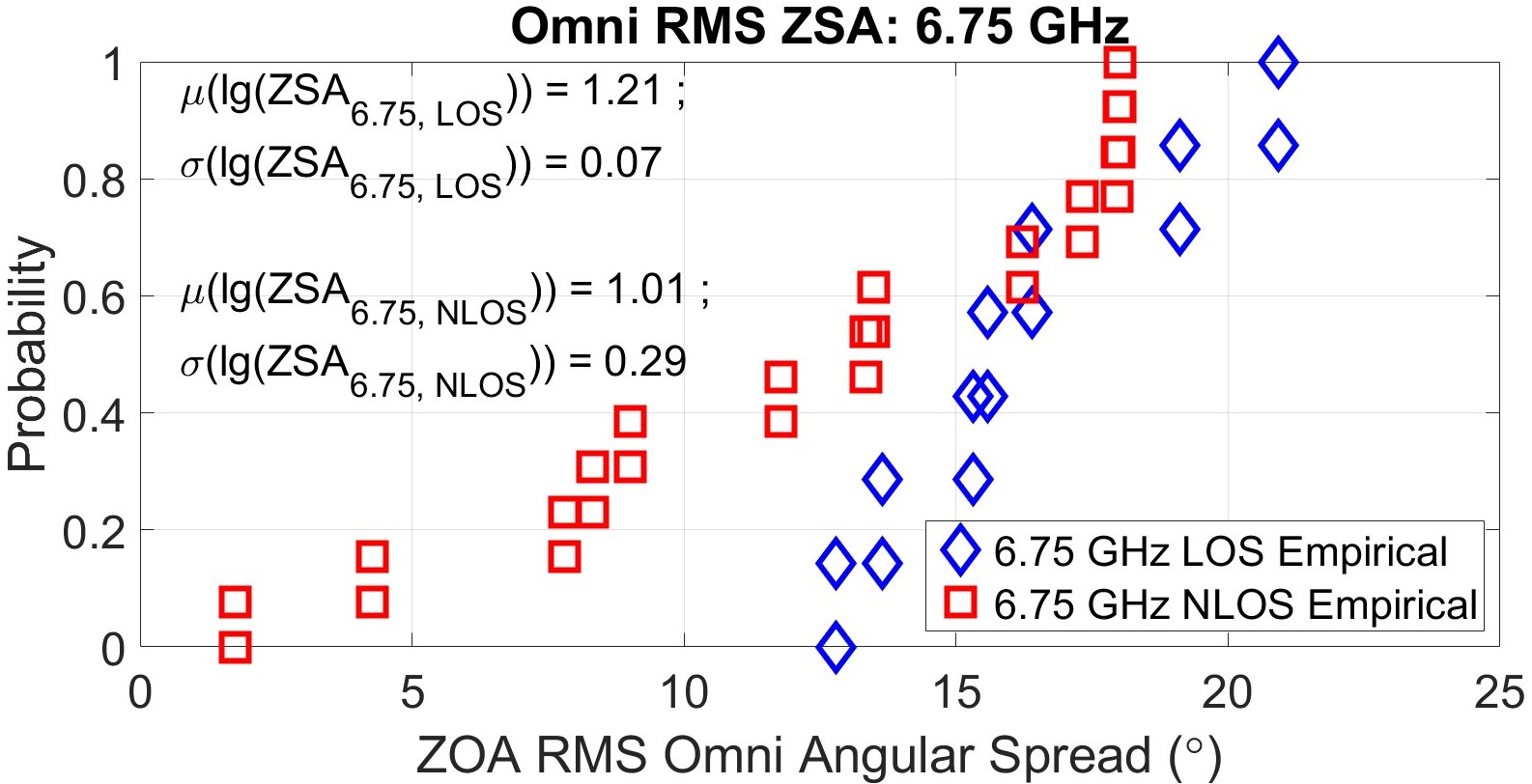}
        \caption{Omni ZSA at 6.75 GHz.}
    \end{subfigure}\\
    \vspace{5pt}
    \begin{subfigure}[t]{0.7\linewidth}
        \centering
        \includegraphics[width=\textwidth]{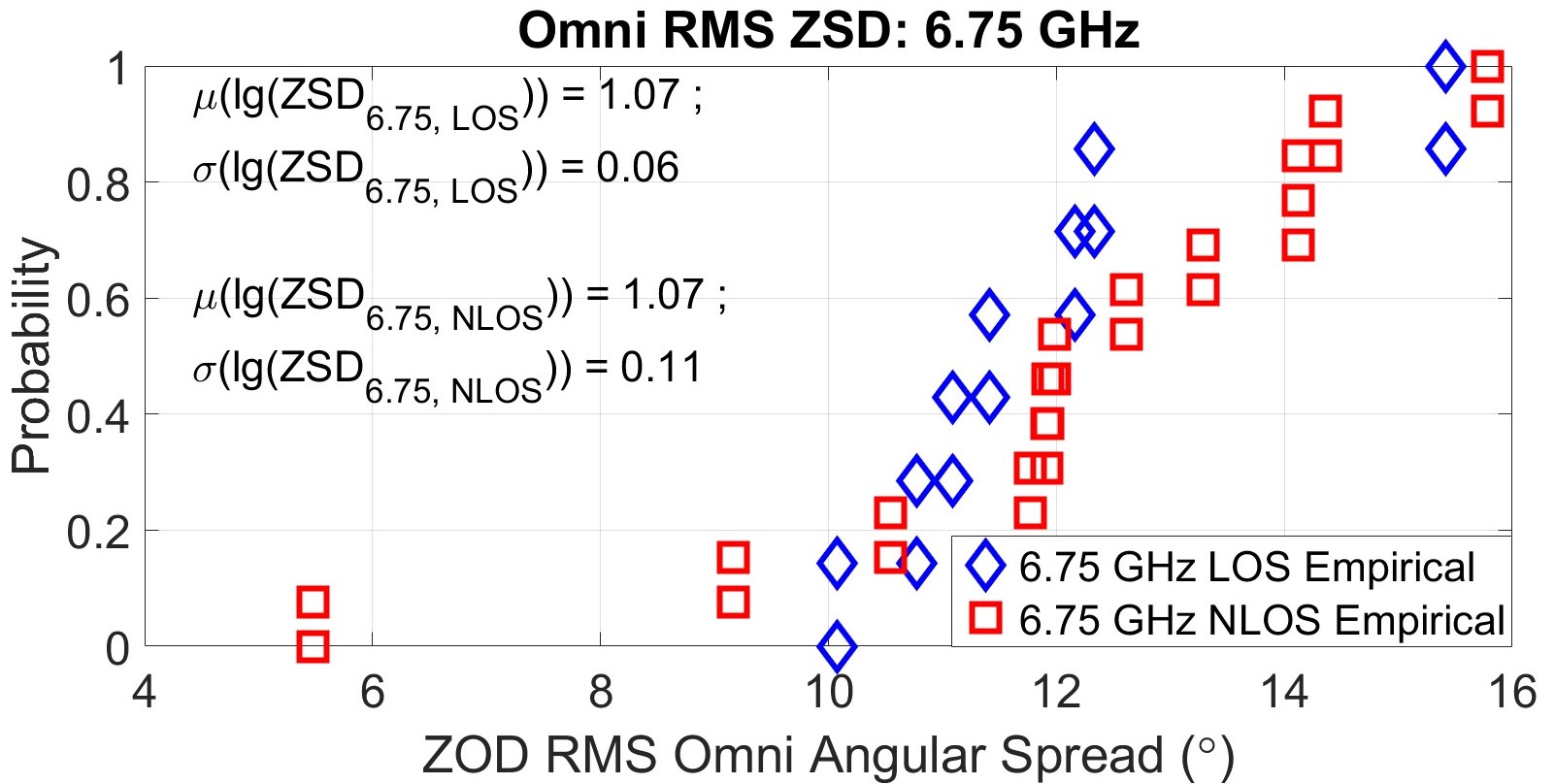}
        \caption{Omni ZSD at 6.75 GHz.}
    \end{subfigure}
    
    \caption{Omnidirectional angular spreads at 6.75 GHz: a)ASA b)ASD c)ZSA d)ZSD.}
    \label{fig:7as}
    \vspace{-20pt}
\end{figure}

\begin{figure}[t]
    \centering
    \begin{subfigure}[t]{0.7\linewidth}
        \centering
        \includegraphics[width=\textwidth]{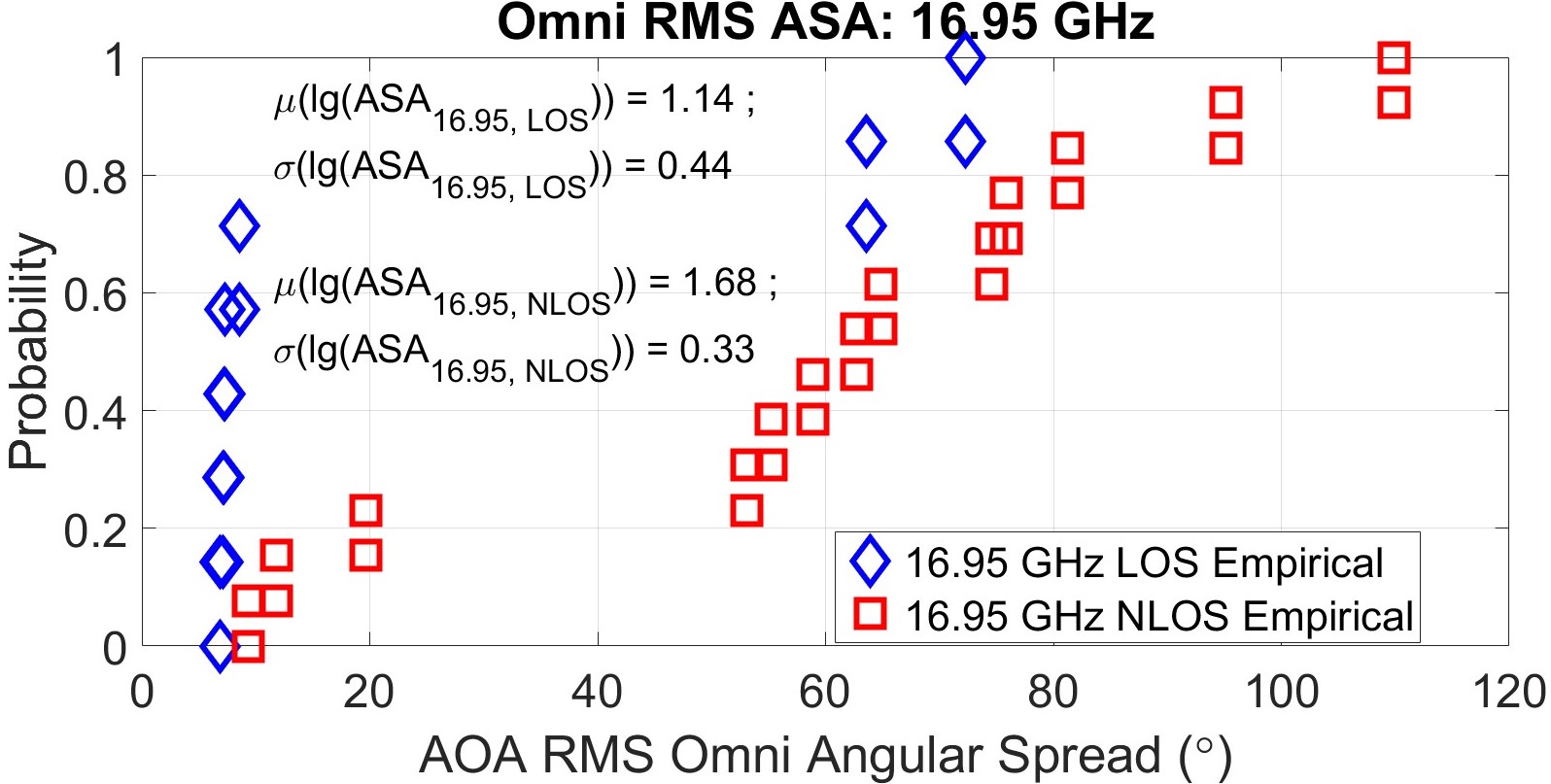}
        \caption{Omni ASA at 16.95 GHz.}
    \end{subfigure}\\
    \vspace{5pt}
    \begin{subfigure}[t]{0.7\linewidth}
        \centering
        \includegraphics[width=\textwidth]{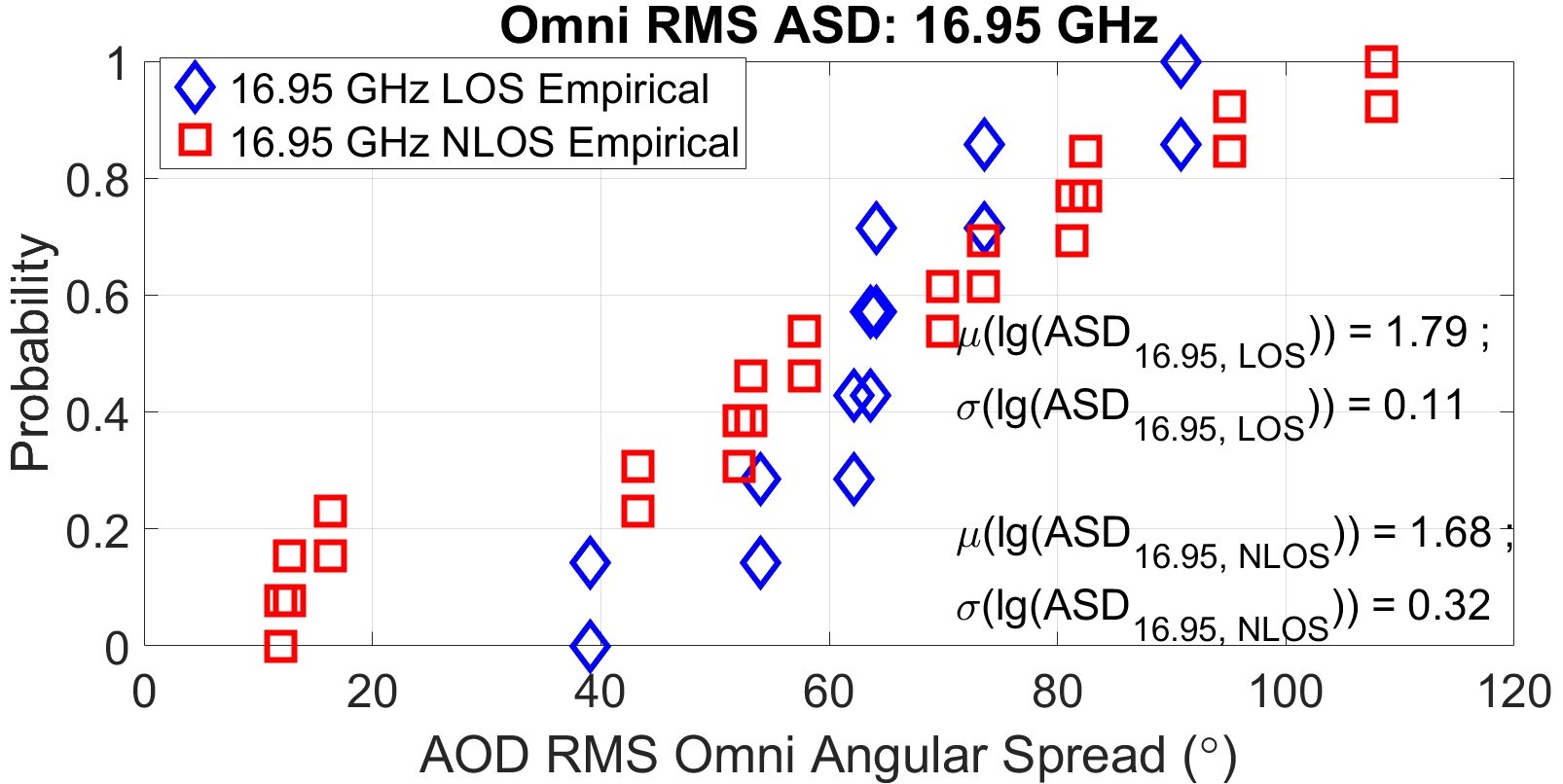}
        \caption{Omni ASD at 16.95 GHz.}
    \end{subfigure}\\
    \vspace{5pt}
    \begin{subfigure}[t]{0.7\linewidth}
        \centering
        \includegraphics[width=\textwidth]{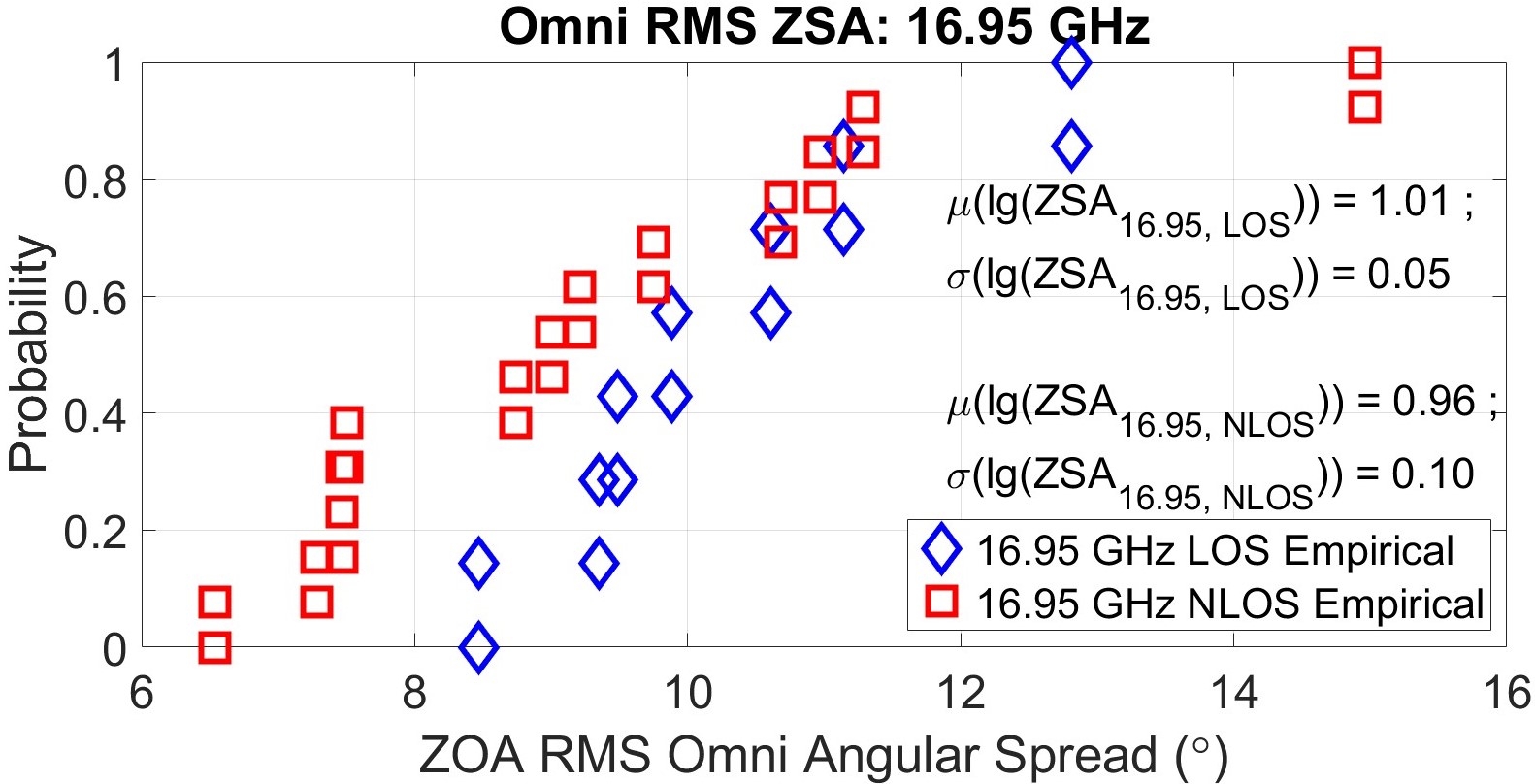}
        \caption{Omni ZSA at 16.95 GHz.}
    \end{subfigure}\\
    \vspace{5pt}
    \begin{subfigure}[t]{0.7\linewidth}
        \centering
        \includegraphics[width=\textwidth]{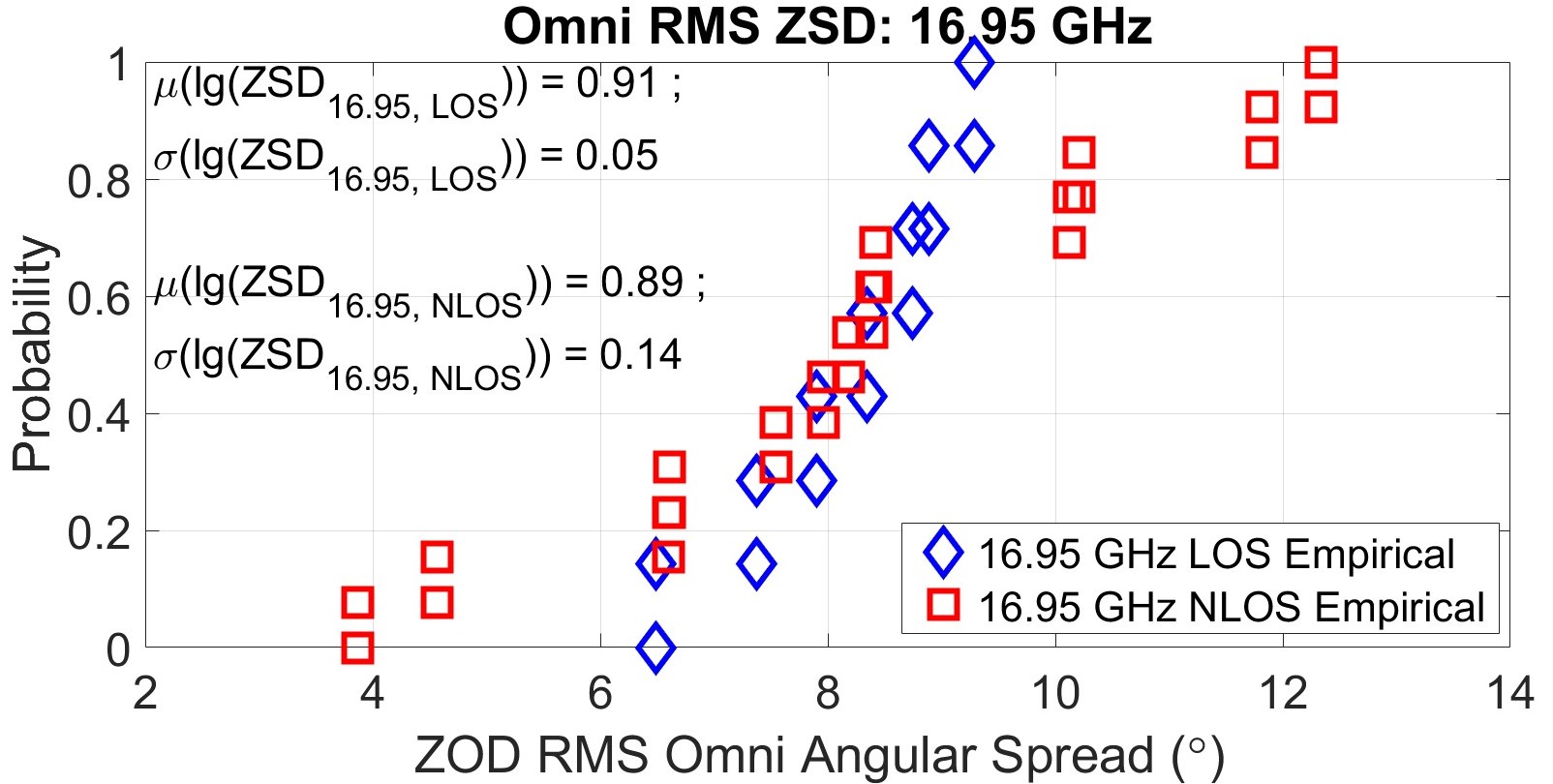}
        \caption{Omni ZSD at 16.95 GHz.}
    \end{subfigure}
    
    \caption{Omnidirectional angular spreads at 16.95 GHz: a)ASA b)ASD c)ZSA d)ZSD.}
    \vspace{-15pt}
    \label{fig:17as}
\end{figure}

\section{Discussions and Comparison with 3GPP models}
\label{diss_comp_3GPP}
\textcolor{black}{The AS measured from the InH campaign showed wider spatial spreads at 6.75 GHz compared to 16.95 GHz in both azimuth and elevation planes. The comparisons in this section were made evaluating the expectation of the underlying log-normal distribution for a physically intuitive comparison in degrees. 
Compared to 3GPP models, NYU ASAs at the RX were similar, with only the NLOS ASA at 16.95 GHz being narrower. NYU ASDs at the TX were wider, while zenith spreads were comparable for LOS, but NYU ZSA and ZSD were narrower than 3GPP values.}

The omni RMS ASA at 6.75 GHz was observed to be 41.3$^{\circ}$ in LOS and 58.1$^{\circ}$ in NLOS, which were slightly smaller than the 3GPP expected values by 1.76$^{\circ}$ and 2.77$^{\circ}$ in LOS and NLOS cases, repectively, as shown in Table \ref{tab:AS}. At 16.95 GHz, the NYU measured omni ASAs were evaluated at 17.2$^{\circ}$ in LOS and 54.3$^{\circ}$ in NLOS. The CDF of AS are presented in Fig. \ref{fig:7as} and \ref{fig:17as}. The 3GPP omni ASAs were obtained as 37.5$^{\circ}$ in LOS and 56.5$^{\circ}$ in NLOS, showing a difference of 17.25$^{\circ}$ in LOS and 2.24$^{\circ}$ in NLOS. Only the LOS ASA at 16.95 GHz was observed smaller for the NYU measurements. The mean lobe RMS ASA for SLs was measured at 10.75$^{\circ}$ in LOS and 28.98$^{\circ}$ in NLOS at 6.75 GHz, and 7.1$^{\circ}$ in LOS and 11.86$^{\circ}$ in NLOS at 16.95 GHz. 


In terms of omni ASD, at 6.75 GHz, 70.3$^{\circ}$ spread was evaluated in LOS, while 48.8$^{\circ}$ in NLOS. At 16.95 GHz, omni RMS ASD was obtained as 62.5$^{\circ}$ in LOS and 53.8$^{\circ}$ in NLOS. The LOS omni ASD was observed wider than the NLOS omni ASD for both 6.75 GHz and 16.95 GHz, which indicated multipath-rich propagation over several different transmit angles in LOS alongside the direct LOS path. The NYU measurements for omni ASD were consistently observed wider than the 3GPP values. At 6.75 GHz, the 3GPP LOS omni ASD of 41.32$^{\circ}$ was 29$^{\circ}$ narrower than the NYU value. In NLOS, the NYU omni ASD was 5.3$^{\circ}$ wider than the 3GPP value of 41.7$^{\circ}$. At 16.95 GHz, the NYU omni ASD was 62.5$^{\circ}$ in LOS and 53.8$^{\circ}$ in NLOS, while the 3GPP model resulted 41.3$^{\circ}$ and 44.8$^{\circ}$ in LOS and NLOS, respectively. 3GPP has modeled the omni ASD as a constant value across frequency, which may need revision. Further, wider measured NYU ASDs, particularly in LOS, compared to 3GPP values at both 6.75 GHz and 16.95 GHz suggest greater diversity in the TX AODs and spatial richness of MPCs.

The mean lobe ASD at 6.75 GHz was observed at 9.8$^{\circ}$ in LOS and 12.7$^{\circ}$ in NLOS. At 16.95 GHz, LOS lobe ASD was observed at 6.2$^{\circ}$, and NLOS lobe ASD was 10.8$^{\circ}$, which were 3.6$^{\circ}$ and 1.9$^{\circ}$ narrower than the lobe ASDs at 6.75 GHz. 

In the zenith plane, the lobe RMS ZSA at 6.75 GHz was observed as 5.7$^{\circ}$ in LOS and 4.5$^{\circ}$ in NLOS, and at 16.95 GHz it was 6$^{\circ}$ in LOS and 5.1$^{\circ}$ in NLOS. Considering omni ZSAs, at 6.75 GHz the ZSA was 16.3$^{\circ}$ in LOS and 11.3$^{\circ}$ in NLOS, and at 16.95 GHz the spread is 10.3$^{\circ}$ in LOS and 9.2$^{\circ}$ in NLOS. CDF plots for zenith spreads are shown in \ref{fig:7as} and \ref{fig:17as}. At 6.75 GHz, 3GPP models resulted in 18.4$^{\circ}$ in LOS, and 29.8$^{\circ}$ in NLOS. At 16.95 GHz, 3GPP RMS ZSA values are obtained as 14.5$^{\circ}$ in LOS and 25$^{\circ}$ in NLOS. 
In NLOS, the NYU-measured ZSAs were smaller than 3GPP ZSAs by 18.5$^{\circ}$ at 6.75 GHz and 15.8$^{\circ}$ at 16.95 GHz. 

The measured ZSD in the spatial lobes were observed as 4.5$^{\circ}$ in LOS and 5.7$^{\circ}$ in NLOS at 6.75 GHz, and 4.8$^{\circ}$ in LOS and 5.4$^{\circ}$ in NLOS at 16.95 GHz. Considering all MPCs and evaluating the omni ZSD, in LOS the spread was found to be 11.8$^{\circ}$ and 11.9$^{\circ}$ in NLOS at 6.75 GHz, while at 16.95 GHz the omni ZSD is evaluated to be 8.2$^{\circ}$ and 7.9$^{\circ}$ in LOS and NLOS, respectively. The 3GPP models yielded 11.2$^{\circ}$ for LOS and 14$^{\circ}$ for NLOS at 6.75 GHz, and 3.4$^{\circ}$ for LOS and 14$^{\circ}$ for NLOS at 16.95 GHz.

Individual MPC azimuth and elevations can be further extracted from the captured PDPs through the use of post-processing methods such as ADME\cite{Ju2024twc} or SAGE\cite{Fleury1999jsac} for a more precise evaluation.

\section{Conclusion}
This paper presented detailed large-scale spatial statistics of wireless channels at 6.75 GHz and 16.95 GHz for InH environments extracted from comprehensive measurements conducted at the NYU WIRELESS Research Center. \textcolor{black}{The AS statistics obtained were compared with the industry standard 3GPP models. The LOS ASA, NLOS ASA, and NLOS ASD were mostly found to be in close agreement with 3GPP models. The NYU-measured LOS ASA at 16.95 GHz was found to be smaller by about 20$^{\circ}$. The measured LOS ASD was found to be larger than the 3GPP model result by about 25$^{\circ}$ on average, indicating a greater diversity in the TX AODs in the azimuth plane. Moreover, in the zenith plane, LOS and NLOS ZSDs and LOS ZSAs were observed to be similar to the 3GPP modeled values, while the NLOS ZSAs were observed to be narrower by about 16$^{\circ}$ for NYU values.} The wide angular spreads clearly indicate spatial richness of MPCs and potential for implementation of spatial multiplexing techniques with MIMO systems in the upper mid-band.

\bibliographystyle{IEEEtran}
\bibliography{references}

\end{document}